\pdfoutput=1

\documentclass[12pt, letter-paper]{article}

\usepackage[utf8]{inputenc}
\usepackage{float}
\usepackage{amsmath}	
\usepackage{amssymb}	
\usepackage{graphicx,psfrag,epsf}
\usepackage{enumerate}
\usepackage{natbib}
\usepackage{url} 
\usepackage{graphicx,color}
\usepackage{hyperref}
\usepackage{amsmath}
\usepackage{ae,aecompl}
\usepackage{courier}
\usepackage{bm}
\usepackage{xcolor}

\usepackage{sectsty}
\sectionfont{\centering}
\usepackage{titlesec}
\titlelabel{\thetitle.\quad}

\newcommand{\blind}{1}

\begin{document}

\def\spacingset#1{\renewcommand{\baselinestretch}%
{#1}\small\normalsize} \spacingset{1}
\newcommand{\argmax}{\mathop{\rm arg~max}\limits}


\if1\blind
{
  \title{\bf Estimation of mask effectiveness perception for small domains using multiple data sources}
  \author{Aditi Sen and Partha Lahiri \thanks{
Aditi Sen is a PhD student, Applied Mathematics \& Statistics, and Scientific Computation, University of Maryland, College Park (e-mail: asen123@umd.edu). Partha Lahiri is Director and Professor, The Joint Program in Survey Methodology \& Department of Mathematics, University of Maryland, College Park, MD  20742, US (e-mail : plahiri@umd.edu). The second author's research was partially supported by  the U.S. National Science Foundation Grant SES-1758808.}\hspace{.2cm}}
   \maketitle
} \fi

\if0\blind
{
  \bigskip
  \bigskip
  \bigskip
  \begin{center}
    {\LARGE\bf Small-area estimation of perception of mask effectiveness for states of the USA}
\end{center}
  \medskip
} \fi

\bigskip

\begin{abstract}
All pandemics are local; so learning  about  the  impacts of pandemics on  public  health  and related societal issues at granular levels is of great interest.  COVID-19 is affecting everyone in the globe and mask wearing is one of the few precautions against it. To quantify people’s perception of mask effectiveness and to prevent the spread of COVID-19 for small areas, we use Understanding America Study’s (UAS) survey data on COVID-19 as our primary data source. Our data analysis shows that direct survey-weighted estimates for small areas could be highly unreliable. In this paper, we develop a synthetic estimation method to estimate proportions of mask effectiveness for small area using a logistic model that combines information from multiple data sources.  We select our working model using an extensive data analysis facilitated by a new variable selection criterion for survey data and benchmarking ratios.  We propose a jackknife method to estimate variance of our proposed estimator. From our data analysis, it is evident that our proposed synthetic method outperforms direct survey-weighted estimator with respect to commonly used evaluation measures.
\end{abstract}

\noindent%
{\it Keywords:} 
Cross-validation; Jackknife; Survey data; Synthetic estimation.
\vfill

\newpage
\spacingset{1.45} 

\section{INTRODUCTION}
\label{sec:intro}
A novel coronavirus has created worldwide crisis in 2020.  \cite{CascellaEtAl2020}  state that published literature can trace the beginning of symptomatic individuals in China back to the beginning of December 2019 and as they were unable to identify the causative agent, these first cases were classified as ``pneumonia of unknown etiology." This new virus is very contagious and has quickly spread globally. The etiology of this illness was consequently attributed to a novel virus belonging to the coronavirus (CoV) family. In a meeting on January 30, 2020, the outbreak was declared by the World Health Organization (WHO) a public health emergency of international concern as it had spread to eighteen countries with four countries reporting human-to-human transmission. The disease the  coronavirus causes was named Covid-19 by the World Health Organization on  February 11, 2020 as depicted in WHO situation reports and activity logs, papers and articles such as \cite{WHOCovidResponse}. The first case of Covid-19 in the USA emerged on January 20, 2020. \cite{HolshueEtAl2020} discuss that the infected was a 35-year-old man who returned to Washington state on January 15, 2020 after traveling to visit family in Wuhan.

The symptoms of Covid-19 range from uncomplicated (mild) illness to moderate pneumonia to severe pneumonia. \cite{CascellaEtAl2020} discuss that for mild cases patients usually present with symptoms of an upper respiratory tract viral infection, including mild fever, cough (dry), sore throat, nasal congestion, malaise, headache, muscle pain, or malaise. New loss of taste and/or smell, diarrhoea, and vomiting are usually observed. Moderate cases show respiratory symptoms such as cough and shortness of breath without signs of severe pneumonia whereas severe pneumonia depicts fever associated with severe dyspnea, respiratory distress, tachypnea, and hypoxia, however, the fever symptom must be interpreted carefully as even in severe forms of the disease, it can be moderate or even absent. The authors further state that based on data from the first cases in Wuhan and investigations conducted by the China CDC and local CDCs, the incubation time could be generally within 3 to 7 days and up to 2 weeks as the longest time from infection to symptoms was 12.5. This data also showed that this novel epidemic doubled about every seven days, whereas the basic reproduction number is 2.2, i.e. on average, each patient transmits the infection to an additional 2.2 individuals. 

Since early months of 2020 every country has experienced the impact of the pandemic in various ways, affecting physical as well as mental and economic health. \cite{PfefferbaumEtAl2020} discuss that the public health emergencies may affect the health, safety, and well-being of both individuals and communities. These effects may translate into a range of emotional reactions. Extensive research in disaster mental health has established that emotional distress is ubiquitous in affected populations. Due to the pandemic life and livelihood has changed manifold with people losing jobs or living with fear of losing income, alienated from society with months of staying at home and as a result of these there is increase in crimes. Referencing \cite{PfefferbaumEtAl2020}, from April 2020, we get to learn that a review of psychological sequelae in samples of quarantined people and of health care providers may be revealed numerous emotional outcomes, including stress, depression, irritability, insomnia, fear, confusion, anger, frustration, boredom, and stigma associated with quarantine, some of which persisted after the quarantine was lifted. Specific stressors included greater duration of confinement, having inadequate supplies, difficulty securing medical care and medications, and resulting financial losses.

As of November 30, 2020 there have been more than 62 million cases of which close to 18 million are active. In terms of the magnitude of infection and deaths the USA remains the country with the most confirmed cases of Covid-19, as of November end the total infection count has surpassed 13 million, obtained from worldometer information: \cite{Worldometer}. In mid April the death toll for USA reached highest in the world, surpassing Italy; on April 11, total death count became greater than 20 thousand in USA, as depicted by various reports like \cite{BBCNewsApril2020}. Through time, testing rates have improved and the Covid Tracking report provides this through charts, \cite{CovidTracking2020}, and daily numbers on hospitalization, recovery, death rates along with testing for various states of the US. USA being one of the most affected countries, is in discussion and an article from the Pew Research Center Social \& Demographic Trends from March 2020, \cite{ PewMAr2020}, note that nearly nine-in-ten U.S. adults say their life has changed at least a little as a result of the COVID-19 outbreak, including 44\% who say their life has changed in a major way. It is further noted that about nine-in-ten U.S. adults (91\%) say that, given the current situation, they would feel uncomfortable attending a crowded party. Roughly three-quarters (77\%) would not want to eat out at a restaurant. In the midst of a presidential election year, about two-thirds (66\%) say they wouldn’t feel comfortable going to a polling place to vote. And smaller but still substantial shares express discomfort even with going to the grocery store (42\%) or visiting with a close friend or family member in their home (38\%).

Response to the pandemic in the US varied from state to state and was famously characterized by an explosion of cases in the state of New York before lockdown was imposed. \cite{GershmanWallStreet2020} discuss that most U.S. states have imposed lockdown measures restricting gathering and social contact, disrupting the lives of hundreds of millions of people and the operations of thousands of businesses. Some states, however, have announced or instituted plans to relax restrictions and several states did not impose lockdown and internal travel was generally unrestricted. Along with quarantine policies the other preventive measures adopted to fight the virus are sanitization and regular hand-washing, wearing masks or face covering while going out in public in order to reduce potential spread of the virus without causing any decelerating impact on the economy like that of lockdown. \cite{GunerEtAl2020} emphasize that with increased testing capacity, detecting more positive patients in the community will also enable the reduction of secondary cases with stricter quarantine rules, but in COVID-19, which has no approved treatment, it is very important to prevent the spread in the society and the main points in preventing the spread in society are hand hygiene, social distancing and quarantine. Earlier due to lack of clarity on the severity of Covid-19 (viz. how fast it spreads or how it can be asymptomatic) some public health officials had suggested it was not mandatory to wear masks or face coverings, but with further development of cases, organizations like the WHO and CDC have suggested it as an effective measure for both people who are affected to stop spreading of the virus and those around not getting affected from contact with affected people: WHO Interim Guidance in January 2020, \cite{WH029Jan2020}, state that ``Wearing a medical mask is one of the prevention measures to limit spread of certain respiratory diseases, including 2019-nCoV, in affected areas. However, the use of a mask alone is insufficient to provide the adequate level of protection and other equally relevant measures should be adopted. If masks are to be used, this measure must be combined with hand hygiene and other IPC measures to prevent the human-tohuman transmission of 2019-nCov."

The issue of wearing mask or face coverings has created a lot of controversy due to difference in opinion and has been highly politicized too, especially in the USA. Mask usage and effectiveness has also been studied through surveys on COVID-19. \cite{KnotekEtAl2020} comment that variation is seen in perception of mask effectiveness due to factors like age saying while most respondents indicated that they were extremely likely to wear a mask if required by public authorities, the reported likelihood is strongly dependent on age and perceived mask efficacy i.e. young aged people not considering masks to be that effective as older people. There is also denial from a section of general public as they find it a breach to their freedom if compelled to mask up. Due to mixed messages even from highest authority there has been difference in approach for different states as to whether mask wearing is so effective as to make it a rule to wear them when social distancing is not as much effective. In a vast nation like the U.S. the approach of states to this issue has varied with some states having made it a mandate to wear masks or face coverings, like that of California by requirements of the state’s Department of Public Health released in \cite{CaliforniaMaskGuidanceJun2020}, which state that ``people who are infected but are asymptomatic or presymptomatic play an important part in community spread. The use of face coverings by everyone can limit the release of infected droplets when talking, coughing, and/or sneezing, as well as reinforce physical distancing".

Understanding America Study, described in section \ref{sec2.1}, is a panel of households at the University of Southern California (USC) of approximately 9,000 respondents representing the entire United States. This survey has been live from March 2020.  Till the time data analysis for this paper is conducted there have been 16 waves. The survey asks pertinent questions related to a variety of topics from COVID-19 Symptoms, Testing, and Medical Care, COVID-19 knowledge, expectations and behaviors, COVID-19 risk perceptions, mental Health and substance abuse, discrimination and stigma, economic and food Security, social safety nets, housing and debt, crime and safety. While for the whole country of USA, national estimates can be effectively derived by weighted means or proportions from respondent level data using relevant variables, but to draw conclusion on small areas such as the different state, for which populations vary a great deal and hence sample size from states vary as well, direct methods of estimation as inappropriate as well as misleading with very low or high estimates and highly variable standard errors. 
 
In this paper, we explore synthetic estimation of the perception on mask effectiveness, i.e., proportion of people considering mask to be highly effective at state level. There is a widespread use of synthetic estimation in different small area aplications; see, e.g., \cite{Rao2015}, \cite{Ghosh20}, 
\cite{HansenEtAl1953}, \cite{Marker1995}, 
\cite{StasnyEtAl1991}, 
and others. Synthetic method uses explicit or implicit models to link several disparate databases in producing efficient estimates for small areas. Along with UAS data, we combine census data, Covid Tracking Report data to create features for models and draw inference on the small areas. The small states of USA like Rhode Island, Wyoming hence can be well represented by synthetics estimates as can be the large states like California, New York.
 
In section 2, we describe primary and supplementary data used in this paper.  In section 3, we evaluate performances of the state level direct survey-weighted estimates.  The performance of the direct method is poor, which motivates synthetic estimation, described in section 4.  In this section, we introduce a jackknife method to estimate  variance of the synthetic estimator. We report main results from our data analysis in section 5. In this section, we introduce a new model selection section criterion for complex survey data in selecting the best performing model.  Finally, we evaluate synthetic estimates by comparative analogy of plotting with direct estimates for a handful of states, some small like District of Columbia, Rhode Island, North Dakota and large states like New York, California, Florida. We conclude the paper with the effectiveness of the methods described in the paper and how they can be extended to any other binary, categorical or continuous variable from this survey or any other with little adjustments or modifications.

\section{DATA USED}
For this study, we will use primary data containing study variable on the perception of mask effectiveness and supplementary data containing  information for building small domain modeling and estimation procedures. 

\subsection{The Primary Data:  Understanding America Study (UAS)} \label{sec2.1}
The Understanding America Study (UAS), conducted by the University of Southern California (USC), is an internet panel of households representing the entire United States. A household is broadly defined as anyone living together with the person who signed up for participating in the UAS.  Using members of the population-representative UAS panel, USC's Center for Economic and Social Research (CESR) launched the Understanding Coronavirus in America tracking survey on March 10.  The survey provides useful information on attitudes, behaviors, including health care avoidance behavior, mental health, personal finances around the novel coronavirus pandemic in the United States.

Initial requests were sent out to the UAS panel members in order to determine their willingness to participate in an ongoing Coronavirus of UAS surveys.  Among 9,063 UAS panel members who responded to the initial request, 8,547  were found eligible to participate in the survey.
On an average till November, 2020 (wave 16) six thousand respondents participate in the surveys, as seen from sample size in Table \ref{tab: wave_details}. Beginning in March 2020, the first round was UAS230, which fielded from March 10 to March 31, 2020, with most responses happening during the period of March 10-14, 2020. UAS 230 is the first round that includes questions that were specifically tailored to COVID-19 and which were repeated in subsequent longitudinal waves. The survey is being conducted in multiple waves. As of November 11, there are 16 waves, as described in Table \ref{tab: wave_details} with their time periods.

\begin{table} 
\caption{Understanding of America Survey (UAS) wave details} 
\label{tab: wave_details} 
\begin{center}
\begin{tabular}{@{\extracolsep{5pt}} rrrrr}
\hline
\hline
    Wave Number & Wave Name & Time period & Sample size \\ 
    \hline
    1 & UAS 230 & March 10,2020 - March 31,2020 & 6,932\\ 
    2 & UAS 235 & April 1, 2020 - April 28, 2020 & 5,478\\
    3 & UAS 240 & April 15, 2020 - May 12, 2020 & 6,287\\
    4 & UAS 242 & April 29, 2020 - May 26, 2020 & 6,403\\
    5 & UAS 244 & May 13, 2020 - June 9, 2020 & 6,407\\
    6 & UAS 246 & May 27, 2020 - June 23, 2020 & 6,408\\
    7 & UAS 248 & June 10, 2020 - July 8, 2020 & 6,346\\ 
    8 & UAS 250 & June 24, 2020 - July 22, 2020 & 6,077\\ 
    9 & UAS 252 & July 8, 2020 - Aug 5, 2020 & 6,289\\ 
    10 & UAS 254 & July 22, 2020 - Aug 19, 2020 & 6,371\\
    11 & UAS 256 & Aug 5, 2020 - Sep 2, 2020 & 6,238\\    
    12 & UAS 258 & Aug 19, 2020 - Sep 16, 2020 & 6,284\\
    13 & UAS 260 & Sep 2, 2020 - Sep 30, 2020 & 6,284\\
    14 & UAS 262 & Sep 16, 2020 - Oct 14, 2020 & 6,129\\
    15 & UAS 264 & Sep 30, 2020 - Oct 27, 2020 & 6,181\\
    16 & UAS 266 & Oct 14, 2020 - Nov 11, 2020 & 6,181\\
\hline
    \end{tabular}
\end{center}
\end{table}

For each wave, eligible panel members are randomly assigned to respond on a specific day so that a full sample is invited to participate over a 14-day period.  Respondents have 14 days to complete the survey but receive an extra monetary incentive for completing the survey on the day they are invited to participate. Thus, except for the first wave, the data collection period for each wave is four weeks with a two-week overlap between any two consecutive waves. 
Each wave data consists of, on an average, six thousand observations.

The UAS is sampled in batches, 
through address-based sampling.  The batches are allocated for national estimation and also for special population estimation (Native Americans, California, and Los Angeles county.)  Essentially UAS is a multiple-frame survey with four frames:  Nationally Representative Sample, Native Americans, Los Angeles (LA) County, and California.   Table \ref{tab: batch_frame_details} shows the relationship between the batches and frames,  but each batch draws from only one frame.

\begin{table}
\caption{Relationship between Batches and Frames in the Understanding America Survey } 
\label{tab: batch_frame_details} 
\begin{center}
\begin{tabular}{@{\extracolsep{5pt}} rr}
\hline
\hline
Batch & Frame\\
\hline
    1 & U.S. \\ 
    2,3 & Native American \\
    4 & Los Angeles County young mothers \\
    5 to 12 & U.S. \\
    13,14,18,19 & Los Angeles County \\
    15,16 & California \\
    17,20,21 & U.S. \\
\hline
    \end{tabular}
\end{center}
\end{table}

As of November 2020, there are 21 batches, the latest being added in August, 2020. Most batches use a two-stage probability sample design in which zip codes are drawn first and then households are drawn at random from the sampled zip codes (except for two small sub-groups which are simple random samples from lists). The National batches draw zip codes without replacement, but the Los Angeles County batches draw with replacement and do sometimes contain the same zip code in different batches.

The final weights, provided in the person level file,  are post-stratified weights starting from base weights. The base weights account for the differential probability of sampling a zip-code and an address within it.  The base weights are then adjusted for nonresponse.  Finally, at the national level, the distribution of nonresponse adjusted weights is calibrated to that of the 2018 Current Population Survey (CPS) weights with respect to selected demographic variables. Weights are provided for all batches, except batch 4, which comprises of Los Angeles County young mothers, and non-Native American households in batches 2 and 3.  \cite{Angrisani2019} describe the sampling and weighting for UAS in great detail.

The survey includes a national bi-weekly long-form questionnaire and a weekly Los Angeles County short-form questionnaire administered in each bi-weekly wave. 
The survey data contains information on different demographic variables such as age, race, sex, and Hispanic origin, education, marital status, work status, identifiers for the states and zip-codes, and various outcome variables affecting human lives (e.g., mental stress, personal finances, COVID-19 like symptoms, testing results, etc.)   The data also contains base and final weights so survey-weighted direct estimates for different outcome variables of interest can be produced.

\subsection{Supplementary Data}

\noindent{\bf The COVID Tracking Project}: Both national and state level data can be downloaded from \url{https://covidtracking.com/}. We use the data as a source of state specific auxiliary variables in our models. The COVID Tracking Project collects and publishes testing data daily for the United States as a whole and also for states and territories. From this data we get to know that for 50 states and DC combined the total test count has been increasing fast with more than 1 million in April to close to total 200 million by end of November. The daily test count also increased from around 180k in April to 1.5 million in November.
There are various state specific auxiliary variables that could be potentially predictive of the perception on  mask effectiveness.  They include COVID-19 daily total testing, total test results, positive/negative, confirmed, death, recovery count (as obtained from Johns Hopkins data on coronavirus), hospitalization, ventilation etc. For this study, our empirical investigation reveals that the following auxiliary variables could be useful in explaining our outcome variable on perception of mask effectiveness:
    \begin{description}
        \item[(i)] totalTestResults: gives total number of tests with positive or negative results,
        \item[(ii)] positive: gives total number of positive tests.
    \end{description}
To make the above two auxiliary variables comparable across 50 states and the District of Columbia, we have used appropriate scaling factors to create the following two auxiliary variables, which we have used in our modeling:

    \begin{description}
        \item[(i)] Testing rate: Total tests with positive or negative results/Total population of state,
        \item[(ii)] Positivity rate: Total positive tests/ Total tests with positive or negative results.
    \end{description}

\vskip .2in
\noindent{\bf Population density data}:    
We use population density estimates in our modeling.  Population density estimates for US states in 2010 are obtained from the US Census  \cite{CensusBureau}
For this study, we have created a categorical feature from it with three levels indicating low (for eg. North Dakota, Wyoming, Alaska etc.), medium(for eg. Georgia, Michigan, Virginia) and high population density(for eg. New York, California, DC) for all states of the USA and DC.

\vskip .2in
\noindent{\bf Democratic party affiliation}:   
According to the modern political party system in the United States, which is a two-party system dominated by the Democratic Party and the Republican Party, we have marked each state as 1 if the political party strength is Democratic in each statewide elective office or 0 if Republican. The information is prior to the 2020 election and obtained from \cite{PoliticalPartyStrength} .

\vskip .2in
\noindent{\bf Region membership of the states}:
Since 1950, the United States Census Bureau defines four statistical regions, with nine divisions. Following information obtained from \cite{ListRegionUS} we have marked each state from each of the four regions - Northeast, Midwest, South and West.

\vskip .2in
\noindent{\bf Census Bureau's Population Estimates Program (PEP)}: For our synthetic estimation method, we need population counts for different demographic groups in the 50 states and the District of Columbia. The Census Bureau releases various tables of population estimates.  On June 2020, the Population Division of the U.S. Census Bureau released annual state resident population estimates by age, sex, race, and Hispanic origin for the period April 1, 2010 to July 1, 2019. 
The Census Bureau essentially obtains these estimates using 2010 decennial census as the base and updates by births, deaths, migration etc. available from the administrative records and others obtained from the ACS survey. We have used two
data sources as follows:
    \begin{enumerate}
        \item SCPRC-EST2019-18+POP-RES: Estimates of the Resident Population Age 18 Years and Older for the US states from July 1, 2019 (released on Dec 2019) which can be directly used.
        \item SC-EST2019-ALLDATA5: Estimates of population by Age, Sex, Race, and Hispanic Origin -- 5 race groups (5 race alone or in combination groups). This data need to be adjusted by filtering out 18+ population (with “AGE”) for the above-mentioned domains (using variables “RACE” for white and rest as other race and “ORIGIN” for Hispanic or Non-Hispanic). Sex is not used, although present in the data and hence set to value 0 for all. The domain wise populations are then adjusted with a factor (i.e. multiplying with domain wise population/total state population) so that the sum of all the domains equalize with the total state level estimate mentioned before.
    \end{enumerate}

\section{DIRECT ESTIMATION}
For the mask effectiveness problem we focus on the following question from survey questionnaire:
    \textit{How effective is wearing a face mask such as the one shown here for keeping you safe from coronavirus? }  This is a categorical variable with five possible answers: \textit{ (i) Extremely Ineffective, (ii) Somewhat Ineffective, (iii)Somewhat Effective, (iv) Extremely Effective, and (v)Unsure.}  The answer choices of respondents have been used to create a binary variable where 1 is taken if mask is considered to be Extremely effective by respondent and 0 otherwise. Using this binary variable the direct estimate works really well at overall national level with low standard error.
    
The survey data contains respondents residing in 50 states and DC, but naturally they are not evenly distributed, i.e., for larger states like California or Florida there is sizable volume in the sample of even as high as 2000 respondents and for smaller states like Delaware or Wyoming there is very little representation of even 3 or 4 respondents. In such scenarios, direct survey-weighted estimates are highly misleading. For example, we see for the first three waves 0\% of people in Wyoming think mask is extremely effective, which happens because all the respondents in the sample take the value 0 for binary response variable mask effectiveness. Hence this is not a good method to draw conclusion for the whole population of the states. 

We observe extremely variable standard error (SE) or margin of error (ME). Estimated SE, or equivalently, estimated ME for a state depends on sample size and value of estimated proportion. For states with small sample sizes, say less than 12, SE is either 0 or very high.  From computations of direct estimates or weighted mean, i.e., the Horvitz-Thompson estimator, from multiple waves we see that for Rhode Island, a state whose contribution in the wave is small with 2 or 3 respondents, estimated SE in the first few waves (1 and 2) is 0\%.  The reason for 0 SE can be either a sample size of only 1 respondent or all binary observations taking the same value (either 0 or 1). In this case of Rhode Island the cause is latter.  But as soon as we have a mix of 0s and 1s, SE becomes very high, as high as even 30\% from wave 5 onwards to wave 9 for Rhode Island. 

\begin{figure}
\begin{center}
\begin{minipage}{.4\textwidth}
  \centering
  \includegraphics[width=1\linewidth]{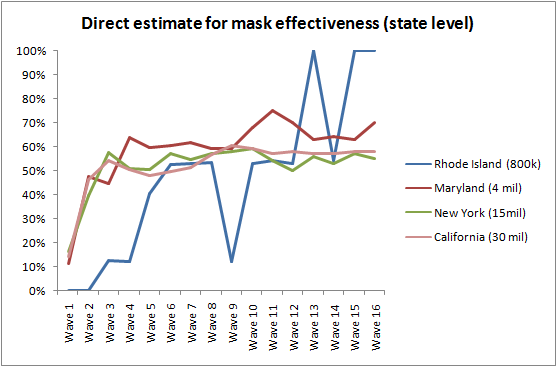}
\end{minipage} \qquad
\begin{minipage}{.4\textwidth}
  \centering
  \includegraphics[width=1\linewidth]{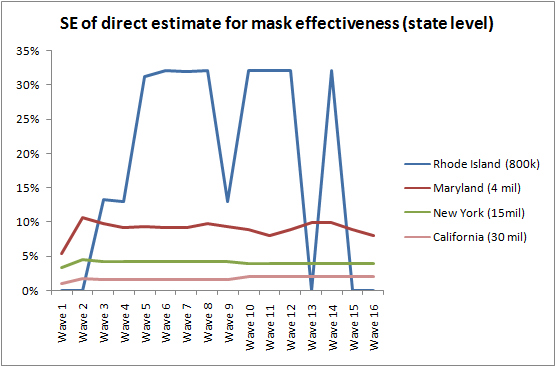}
\end{minipage}
\vskip .2in
\end{center}
\caption{Direct estimates of mask effectiveness and associated standard errors for four selected states.
\label{fig:se_direct_state_estimate}}
\end{figure}

We have showed the erratic behavior of direct estimates  and standard errors as comparative view of four states with varying population sizes (as estimated from the Census Bureau's PEP data)- one with high population (California - estimated adult population of 30 million from PEP), one with medium population (New York - estimated adult population of 15 million from PEP), one with small population (Maryland - estimated adult population of 4 million from PEP) and one with very small population (Rhode Island - estimated adult population of 800k from PEP), in Figure \ref{fig:se_direct_state_estimate}. The curves for Rhode Island are the most variable (both state level direct estimate and SE), next is Maryland, which vary but not too much and those of New York and California, which are quite stable. These SE estimates are thus surely very unstable or unreliable and typically, in public opinion polls like the famous Gallup polls, margin of errors (2SE) is targeted at a low level such as 3\% or 4\%.  
The Figure \ref{fig:se_direct_state_estimate} for Rhode Island demonstrates high variability in state estimates for smaller states.

\begin{figure}
\begin{center}
\begin{minipage}{.4\textwidth}
  \centering
  \includegraphics[width=1\linewidth]{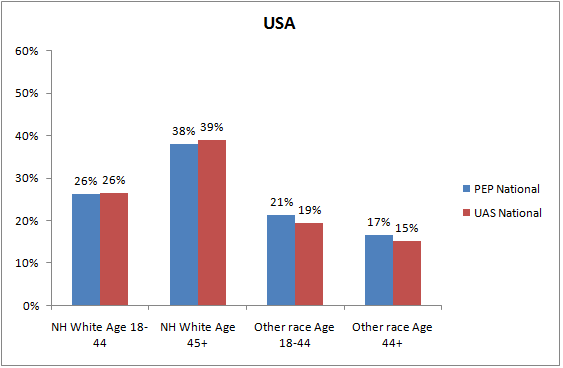}
\end{minipage} \qquad
\begin{minipage}{.4\textwidth}
  \centering
  \includegraphics[width=1\linewidth]{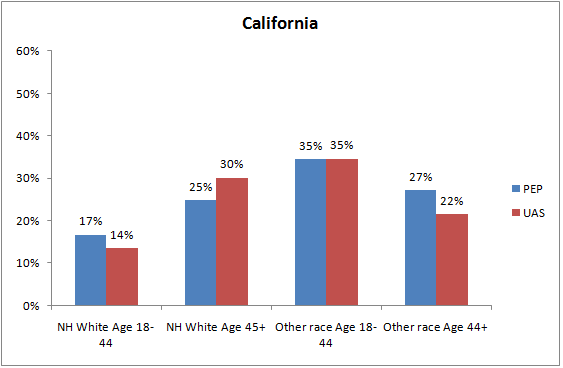}
\end{minipage}
\vskip .2in

\begin{minipage}{.4\textwidth}
\centering
  \includegraphics[width=1\linewidth]{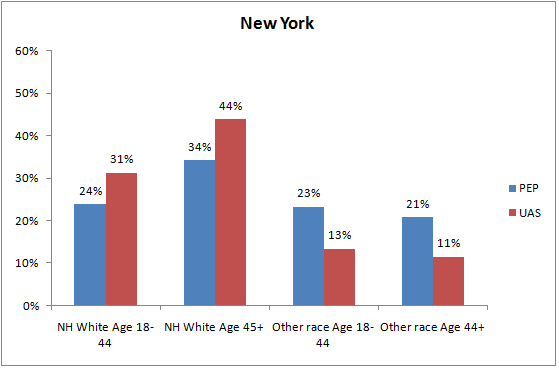}
\end{minipage} \qquad
\begin{minipage}{.4\textwidth}
  \centering
  \includegraphics[width=1\linewidth]{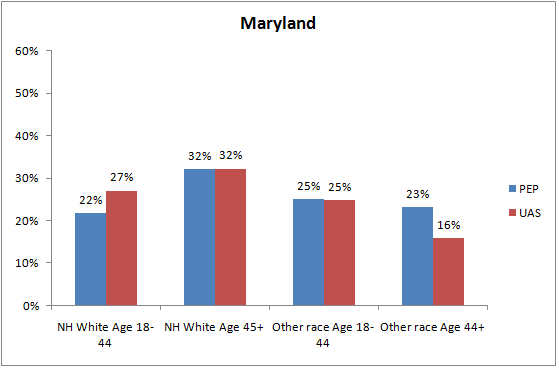}
\end{minipage}
\vskip .2in

\begin{minipage}{.4\textwidth}
  \centering
  \includegraphics[width=1\linewidth]{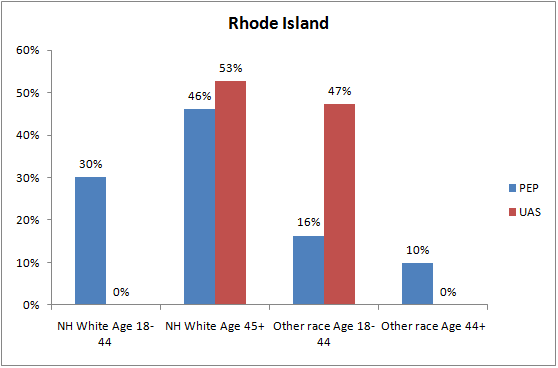}
\end{minipage} \qquad
\begin{minipage}{.4\textwidth}
  \centering
  \includegraphics[width=1\linewidth]{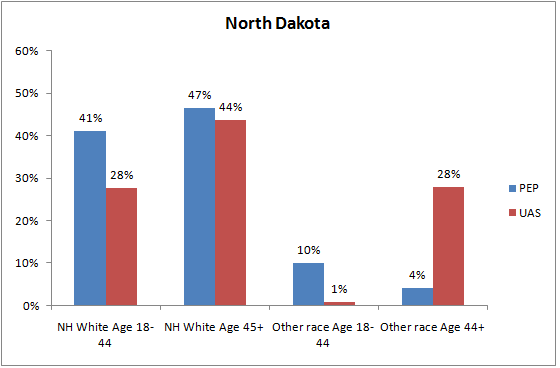}
\end{minipage}
\end{center}
\vskip .2in
\caption{PEP vs. UAS estimates of 4 domains
\label{fig:PEP_UAS_small}}
\end{figure}

Along with high variability a demonstration of high bias in the direct state estimates can also be viewed. Since we do not know the truth for mask effectiveness, we cannot demonstrate bias properties for mask effectiveness. But we can say if we consider another outcome variable for which ``truth" is known from the PEP data, we can at least partially justify our claim. Using Figure \ref{fig:PEP_UAS_small} we show that UAS estimates of proportions of people falling in the four demographic groups or domains we considered do not match up with PEP data for states, but they more or less match at the national level. For large states like California, the difference between PEP estimate of percentage of adult population and UAS direct wave estimate is little, similar it is for medium sized states like Maryland and New York, but for small states like Rhode Island and North Dakota, referring to Figure \ref{fig:PEP_UAS_small}, we see the percentages vary a great with even 0\% or no contribution in some domains.

\section{SYNTHETIC METHOD}

For the synthetic estimation of mask effectiveness for small areas, i.e., at state level we first define the following  notations and then derive the formula for the estimator from a logistic regression model. 
Let $y_{k}$ denote the value of outcome (or dependent) variable for the $k$th respondent ($k=1,\cdots,n$), where $n$ denotes the number of persons in a given wave (say, wave 2 covering April 1-April 28) of the UAS survey. The outcome variable is binary as defined by $y_k=1$ if person $k$ considers mask wearing highly effective. Let $x_k=(x_{k1},\cdots,x_{kp})'$ denote the value of a vector of auxiliary variables (same as independent variables or predictor variables or covariates) for respondent $k$. We have focused on the following two criteria for selecting the auxiliary variables for the unit level logistic regression model:
(i) auxiliary variables should have good explanatory power in explaining the outcome variable of interest; (ii) total or mean of these auxiliary variables for the population should be available from a big data such as a bigger survey, administrative records or decennial census. 
Let $N_i$ and $N_{gi}$ be the population size of the adult (18+) and the $g$th cell in state $i$, respectively. As discussed previously in the data section the $N_{gi}$ and $N_i$ values are obtained from the US Census Bureau. 
Let $y_{gik}$ be the value of the outcome variable for $k$th person in state $i$ for the $g$th group ($g=1,\cdots,G;\;i=1,\cdots,m;\;k=1,\cdots,N_{ig}$).  Here $m=51$ (50 states and DC) are the small areas. Let $z_i$ be a vector of state specific auxiliary variables.
For the estimation of mask-effectiveness variable for the 50 states and the District of Columbia, we write the population model as:
\begin{eqnarray}
\label{popmodel}
\mbox{Level 1:}\;\; y_{gik}|\theta_{gi}\stackrel{ind}\sim f(\theta_{gi}),\;\;\;\;\;
\mbox{Level 1:}\;\; h(\theta_{gi})=x'_g\beta+z'_i\gamma\nonumber,
\end{eqnarray}
for $k=1,\cdots,N_{gi},\;i=1,\cdots, m;\;g=1,\cdots,G$, where $f(\theta_{gi})$ is suitable distribution with parameter $\theta_{gik}$ (here for binary variable this is a Bernoulli distribution with success probability $\theta_{gik}$); $g(\theta_{gik})$ is a suitable known link function (for this application, we take logit link); $\beta$ and $\gamma$ are unknown parameter to be estimated using UAS micro data, i.e., at respondent or unit level using survey weights. 

We estimate population mean for state $i$ by:
$\hat{\bar Y}_i^{syn}=\sum_{g=1}^G (N_{gi}/N_i)\hat\theta_{gi}=\sum_{g=1}^G (N_{gi}/N_i)\\h^{-1}(x'_g\hat\beta+z'_i\hat\gamma),$
where $h^{-1}$ is the inverse function of $h$; $\hat\beta$ and $\hat\gamma$ are the survey-weighted estimator of $\beta$ and $\gamma$, respectively.
If $h(\cdot)$ is a logit function, we have
$\hat{\bar Y}_i^{syn}=\sum_{g=1}^G (N_{gi}/N_i)\hat\theta_{gi}=\sum_{g=1}^G (N_{gi}/N_i)\exp (x'_g\hat\beta+z'_i\hat\gamma)\left [1+\exp (x'_g\hat\beta+z'_i\hat\gamma)\right ]^{-1}.$

We propose a  jackknife method to estimate the variance of the proposed synthetic estimator. We obtain $i$th jackknife resample by deleting all survey observations in batch $j$.  Thus we have $m=20$ jackknife resamples from wave 14 onwards because there are 20 batches in total, whereas earlier for waves 1 to 13 there were in total 19 batches in each wave data, the latest addition being ``21 MSG 2020/08 Nat. Rep. Batch 11" in August and LA County Young mothers is not present in any of the waves. For each jackknife resample, we recompute replicate synthetic estimate given using (\ref{popmodel}).  We will get $m$ such replicate estimates, say, $\hat {\bar Y}_{i(-j)}^{syn}\;(j=1,\cdots,m)$.  We can then estimate the variance of $\hat {\bar Y}_{i}^{syn}$ by
    \begin{eqnarray}\label{synvar}
    v(\hat {\bar Y}_{i}^{syn})=\frac{m-1}{m}\sum_{j=1}^m\left (\hat {\bar Y}_{i(-j)}^{syn}-\frac{1}{m}\sum_{j=1}^m  \hat {\bar Y}_{i(-j)}^{syn} \right )^2.
    \end{eqnarray}

\section{DATA ANALYSIS}

\begin{figure}
\begin{center}
\begin{minipage}{.4\textwidth}
  \includegraphics[width=1\linewidth]{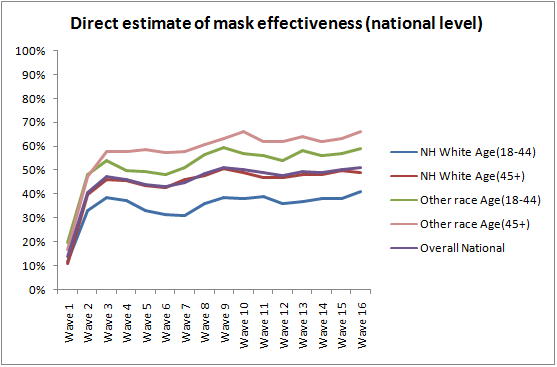}
\end{minipage} \qquad
\begin{minipage}{.4\textwidth}
  \centering
  \includegraphics[width=1\linewidth]{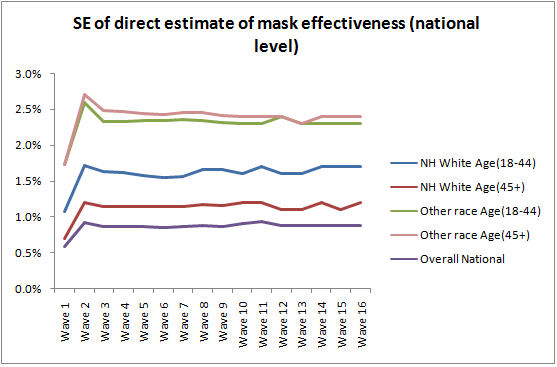}
\end{minipage}
\end{center}
\caption{National direct wave estimates of mask effectiveness and associated standard error direct estimates; overall estimates as well as estimates for four groups are provided.
\label{fig:national_direct_se_groups}}
\end{figure}

At national level in order to understand the broader question on the identification of demographic factors influencing effectiveness perceptions certain domains or groups are created based on race-ethnicity x age. These four groups are Non Hispanic White Age 18-44, Non Hispanic White Age 45+, Other race Age 18-44 and Other race Age 44+.  Considerable variation among these groups is observed across multiple waves with all standard errors (SE) from direct estimates around 2\%, after which it is chosen for further estimation study. The direct survey-weighted estimates at the national level as well as domain level from waves 1 to 9 are provided in Table \ref{tab: national_direct_se_domain_estimates}  along with the standard errors in parenthesis; see also Figure \ref{fig:national_direct_se_groups}. We observe that the overall national estimate and the domain NH White Age(45+) behave similarly (e.g, 50\% and 49\% at wave 10, respectively). The Other Race Age (45+) domain has the highest perception of mask effectiveness (e.g.,  66\% at wave 10), whereas the domain NH White Age (18-44) has the least value of such estimate (e.g., 38\% at wave 10). Thus this breakdown of the population into domains can be used further into the analysis during modelling. We have used R survey package to compute such estimates with the weights of respondents as provided in the wave data. 

\begin{table}
\caption{Direct national estimates of mask effectiveness (associated standard errors) for selected demographic groups and first five waves.
 } 
\label{tab: national_direct_se_domain_estimates} 
\begin{center}
\begin{tabular}{@{\extracolsep{5pt}} rrrrrr}
\hline
\hline
Direct Estimate & Wave 1 & Wave 2 & Wave 3 & Wave 4 & Wave 5\\
\hline
    Overall National & \shortstack{14\% \\ (0.6\%)} & \shortstack{41\% \\ (0.9\%)} & \shortstack{47\% \\ (0.9\%)} & \shortstack{46\% \\ (0.9\%)} & \shortstack{44\% \\ (0.9\%)}\\ 
    NH White Age(18-44) & \shortstack{12\% \\ (1.1\%)} & \shortstack{33\% \\ (1.7\%)} & \shortstack{39\% \\ (1.6\%)} & \shortstack{37\% \\ (1.6\%)} & \shortstack{33\% \\ (1.6\%)}\\ 
    NH White Age(45-all) & \shortstack{11\% \\ (0.7\%)} & \shortstack{40\% \\ (1.2\%)} & \shortstack{46\% \\ (1.1\%)} & \shortstack{46\% \\ (1.1\%)} & \shortstack{44\% \\ (1.1\%)}\\
    Other race Age(18-44) & \shortstack{20\% \\ (1.7\%)} & \shortstack{48\% \\ (2.6\%)} & \shortstack{54\% \\ (2.3\%)} & \shortstack{50\% \\ (2.3\%)} & \shortstack{49\% \\ (2.3\%)}\\
    Other race Age(45-all) & \shortstack{17\% \\ (1.7\%)} & \shortstack{47\% \\ (2.7\%)} & \shortstack{58\% \\ (2.5\%)} & \shortstack{58\% \\ (2.5\%)} & \shortstack{58\% \\ (2.4\%)}\\
\hline
    \end{tabular}
\end{center}
\end{table}

From the aforementioned observations, it is clear the direct estimates are not stable even at the state level. The synthetic estimators essentially would borrow strength from other states through implicit or explicit models and combine information from the sample survey, various administrative/census records, or previous surveys. Synthetic estimators are highly effective and appealing in estimating any response for small areas; in this case for US states the proportion of people considering mask to be highly effective against the spread of coronavirus. Referring to synthetic estimation methods explained in \cite{LahiriPramanik2019}, we employ unit level logistic model with respondent level characteristics like the age x race/ethnicity  along with state level auxiliary variables such as regional identifier (e.g., Northeast, Midwest, South or West), party affiliation of state governor or DC mayor (Democratic or Republic) and even the state level Covid-19 testing or positivity rate. Thus we have combined the data in UAS coronavirus survey  with US Census Bureau data and Covid Tracking Project data to derive state level synthetic estimator of population means and totals for the variable of interest. 

\subsection{Variable  Selection}
For all the 16 waves, we first fit the full model, i.e., the model with all auxiliary variables listed above.  Table \ref{tab: model1_results} displays significant auxiliary variables in all the waves. We then concentrate our focus on the models given in   Table \ref{tab: model_list}.  These are logistic regression models for the indicator response variable mask effectiveness with the different combinations of auxiliary  variables.  
In every case we use R survey package to run weighted logistic regression with quasi-binomial family, where weights are the final post-stratification weights as provided by UAS and design is defined with such weights and no strata or cluster. 

    {\tiny
    \begin{table}
        \caption{Significant covariates in Model 1 for different waves;  from R package : Signif. codes:  0 ‘***’ 0.001 ‘**’ 0.01 ‘*’ 0.05 ‘.’ 0.1 ‘ ’ 1}
    \label{tab: model1_results}
    \begin{center}
    \resizebox{\columnwidth}{!}{%
    \begin{tabular}{@{\extracolsep{5pt}} rrrrrrrrrrrr}
    \hline 
    \hline
    Wave & intercept & \shortstack{ NH White \\ Age(18-44) \\ (indicator)} & \shortstack{ NH White \\ Age(44+) \\ (indicator)} & \shortstack{Other race \\ Age(18-44) \\  (indicator)} & \shortstack{testing \\ rate} & \shortstack{positivity \\ rate} & \shortstack{population \\ density \\ (categorical)} & \shortstack{region \\ Northeast \\ (indicator)} & \shortstack{region \\ Midwest \\ (indicator)} & \shortstack{region \\ South \\ (indicator)} & \shortstack{Democratic \\ party \\ (indicator)} \\
    \hline
    1 & *** & *   & ***  &    &           &    &     &           & $\bullet$ & ** &  \\
    2 &     &***  & *   &    & $\bullet$ &    &     &           &  &           &  \\
    3 &     & *** & *** &    &           &    & ***    &        &  &           &  \\
    4 &     &***  & *** &  * &           &    &***  &           &  & **        &  \\
    5 &*    & *** & *** & *  &           & ** &**   & $\bullet$ &  &           &  \\
    6 &     & *** & *** & *  &           &    & *** &           &  &           & \\
    ~~\\
    7 &     & *** & *** &  *     &           &    & **  &           &  &           &  \\
    8 &     &***  & *** &        &           &    & *** &           &* &          &  \\
    9 &     & *** & *** &        &  $\bullet$&    & *** &           &  &           &  \\
    10&     &***  & *** &  *     &   *       &    &**   &           &  &           &  \\
    11&     & *** & *** &$\bullet$&           &    &***  &           & $\bullet$ &$\bullet$           &  \\
    12 &   & *** & *** & *        &           &    & *** &           &  &           & \\
    ~~\\
    13 &    & *** & *** &$\bullet$& *        &    & ** &           &  &      ** &  \\
    14&     &***  & *** &$\bullet$&          &    &***   &           &  &           &  \\
    15&     & *** & *** &$\bullet$&           &    &  &           & $\bullet$ &           &  \\
    16 &   & *** & *** & *        &           &    & *** &           &  &           & \\
    \hline
    \end{tabular}%
    } 
    \end{center}
    \end{table}
    }

{\tiny
    \begin{table}
     \caption{A list of competing models} 
    \label{tab: model_list}
    \begin{center}
    \resizebox{\columnwidth}{!}{%
    \begin{tabular}{@{\extracolsep{5pt}}  rrrrrrrrrrrr}
    \hline 
    \hline 
    Wave & intercept & \shortstack{ NH White \\ Age(18-44) \\ (indicator)} & \shortstack{ NH White \\ Age(44+) \\ (indicator)} & \shortstack{Other race \\ Age(18-44) \\  (indicator)} & \shortstack{testing \\ rate} & \shortstack{positivity \\ rate} & \shortstack{population \\ density \\ (categorical)} & \shortstack{region \\ Northeast \\ (indicator)} & \shortstack{region \\ Midwest \\ (indicator)} & \shortstack{region \\ South \\ (indicator)} & \shortstack{Democratic \\ party \\ (indicator)} \\
    \hline
    M1 & \checkmark & \checkmark & \checkmark & \checkmark & \checkmark & \checkmark & \checkmark & \checkmark & \checkmark & \checkmark & \checkmark \\
    M2& \checkmark & \checkmark & \checkmark & \checkmark & \checkmark & \checkmark & \checkmark & \checkmark & \checkmark & \checkmark &  \\
    M3& \checkmark & \checkmark & \checkmark & \checkmark & \checkmark &  & \checkmark & \checkmark & \checkmark & \checkmark &  \\
    M4 & \checkmark & \checkmark & \checkmark & \checkmark & \checkmark &  & \checkmark &  & & \checkmark &  \\
    M5 & \checkmark & \checkmark & \checkmark & \checkmark &  &  & \checkmark &  & & \checkmark &  \\
    M6 & \checkmark & \checkmark & \checkmark & \checkmark & \checkmark &  & \checkmark &  & &  &  \\
    M7 & \checkmark & \checkmark & \checkmark & \checkmark &  &  & \checkmark &  & &  &  \\
    
    \hline
    \end{tabular}%
    } 
    \end{center}
    \end{table}
    }
    
     The full model, i.e., M1,  is our starting point. 
     In model selection terminology, this is indeed a correct model.  In other words, true values of some of the coefficients of M1 may be zero;  if the sample size is large, those coefficients will be estimated at near zero.  But, if we keep too many covariates in a model, the estimates  may be subject  to high variability (and thereby we may lose some predictive power if we select model with a lot of covariates.)   
    
    We now explore the possibility of reducing the number of auxiliary variables from M1.  There are a large number of possible models so we proceed systematically.  To this end, we fit M1 for all the 16 waves. Table \ref{tab: model1_results} reports significant auxiliary variables for each of the 16 waves. In all the models, we  include intercept (whether or not it is significant). Using information in Table \ref{tab: model1_results}, we create Table \ref{tab: model_list}, which lists a number of competing models with less number of model parameters. We now explain why we want to consider models M1-M7 for further comparison.  
    
    All the auxiliary variables except for the democratic party affiliation appear in at least one wave. Thus natural question is what happens if we drop the democratic party affiliation from M1, which motivates keeping M2 for further investigation.  Positivity rate is significant only in wave 5.  This suggests inclusion of model M3  for further investigation. The factors NH Whites (18-44), NH Whites (44+), Other Race (18-44), population density are all significant for 5 waves: 6,7, 12, 14, and 16.  So the model M7 seems to be a natural choice.  We then consider models M3-6.  Note that, in addition to NH Whites (18-44), NH Whites (44+), Other Race (18-44), population density, each of these three models includes an additional auxiliary variable significant in at least one wave.  For example, M4 includes additional auxiliary variable testing rate because all M4 coefficients are significant in wave 10; similarly.

   To select one out of the seven models listed in Table \ref{tab: model_list}, we apply a  cross-validation leave-one-state-out method.  We now describe the method. We leave out the entire UAS survey data on the outcome variable $y_i$ (e.g., mask effectiveness) for state $i$ and predict the vector of outcome variables for all sampled units of the leave out state using $x_g$ for the sampled unit and $z_i$ for the leave out state.  Let $f(y_i|y_{-i})$ denote the joint density of $y_i$, all the observations in state $i$, conditional on the data from the rest of the states, say $y_{-i}$.  For the Bernoulli distribution of $y_i$ for state $i$, using independence, we have for known model parameters $\beta$ and $\gamma$:
        \begin{eqnarray*}
        \log f(y_i|y_{-i};\beta,\gamma)&=&\sum_{g=1}^G\sum_{k=1}^{n_{gi}}\left [ y_{gik}\log \theta_{gi}+(n_{gi}-y_{gik})\log (1-\theta_{gi})  \right ]\\
                          &=&\sum_{g=1}^G\sum_{k=1}^{n_{gi}}\left [ y_{gik}\log \left (\frac{\theta_{gi}}{1-\theta_{gi} }\right )+n_{gi}\log (1-\theta_{gi})  \right ]\\
                         &=&\sum_{g=1}^G\sum_{k=1}^{n_{gi}}\left [ y_{gik}(x'_g\beta+z'_i\gamma) - n_{gi}\log \left (1+\exp (x'_g\beta+z'_i\gamma)\right )  \right ].
                        \end{eqnarray*}
        Using data from the rest of states, i.e., $y_{(-i)}$ we get the survey-weighted estimates $\beta$ and $\gamma$ and plug in the above expression.  Let these estimates be $\hat\beta_{w,(-i)}$ and $\hat\gamma_{w;(-i)}$.  We then define our model selection criterion as:
        $$ C= \sum_{i=1}\sum_{g=1}^G\sum_{k=1}^{n_{gi}}w_{gik}\left [ y_{gik}(x'_g\hat\beta_{w,(-i)}+z'_i\hat\gamma_{w;(-i)}) - n_{gi}\log \left (1+\exp (x'_g\hat\beta_{w,(-i)}+z'_i\hat\gamma_{w;(-i)})\right )  \right ]. $$

     For each of the models in Table \ref{tab: model_list}, we compute C model selection measures for all the waves (wave 1-16).  In Table \ref{tab: c_value}, we report the quantiles (minimum, first quartile, median, third quartile, maximum) and mean of C values (over the 16 waves) for each model in Table \ref{tab: model_list}.  We divide C value from each model by the sample size of the wave
    to scale down the numbers for ease of comparison.  The C values are all negative, as probability densities are values between 0 and 1 and logarithm of fractions make these  negative. For every state, iteratively regressions are run and regression estimates are obtained, which are used in the formula. Using Table \ref{tab: c_value}, we select M2 as the best performing model because this model produces maximum value of all descriptive statistics reported in Table \ref{tab: c_value}.

    \begin{table}
    \caption{Cross validation leave one state out statistic for all models
    \label{tab: c_value}}
    \begin{center}
       \begin{tabular}{@{\extracolsep{5pt}}rrrrrrr} 
    \hline
    \hline
        Model & $0\%$  & $25\%$& $50\%$&$75\%$& $100\%$ & Mean \\ 
        \hline
        M1  & -89& -83& -75& -68& -16& -71 \\
        M2  & -79& -55& -10& -8& -7& -28 \\
        M3  & -92& -86& -83& -78& -20& -78 \\
        M4  & -92& -86& -82& -77& -21& -78 \\
        M5  & -91& -84& -80& -76& -19& -76 \\
        M6  & -68& -59& -55& -51& -17& -54 \\
        M7  & -90& -82& -76& -69& -15& -72 \\
       
        \hline
        \end{tabular}
        \end{center}
    \end{table}

    \subsection{\textbf{Synthetic estimation of the perception of mask effectiveness for the states}}
    
\begin{figure}
\centering
\begin{minipage}{.4\textwidth}
  \centering
  \includegraphics[width=1\linewidth]{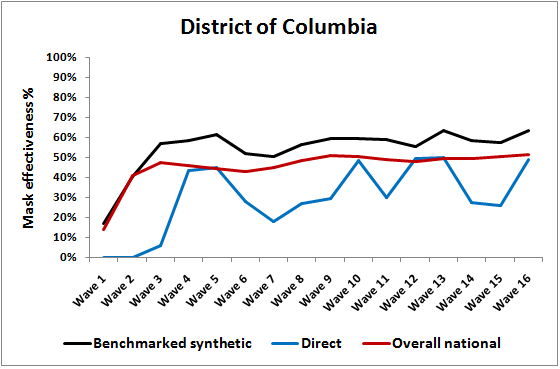}
 \end{minipage} \qquad
\begin{minipage}{.4\textwidth}
  \centering
  \includegraphics[width=1\linewidth]{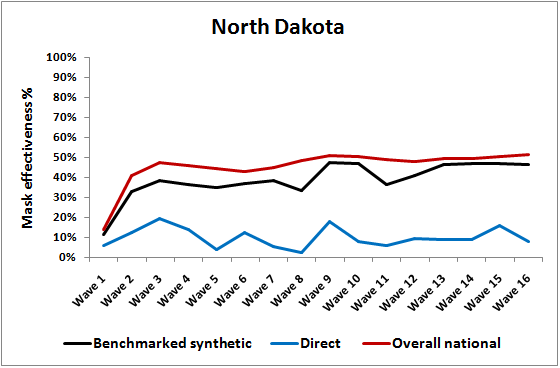}
 \end{minipage}
\vskip .2in
\begin{minipage}{.4\textwidth}
  \centering
  \includegraphics[width=1\linewidth]{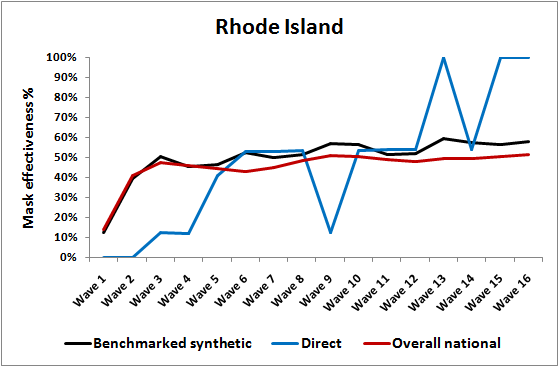}
  \end{minipage} \qquad
\begin{minipage}{.4\textwidth}
  \centering
  \includegraphics[width=1\linewidth]{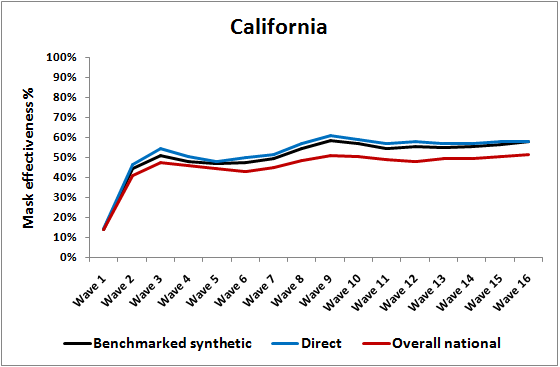}
 \end{minipage}
\vskip .2in
\begin{minipage}{.4\textwidth}
  \centering
  \includegraphics[width=1\linewidth]{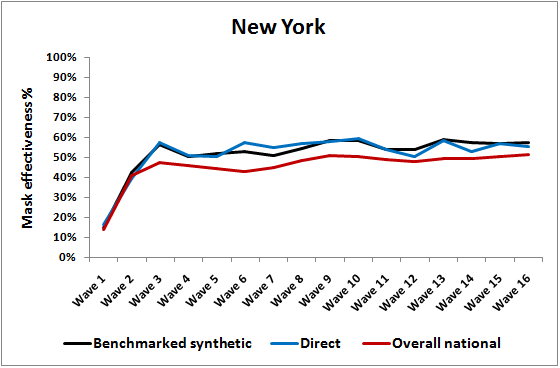}
 \end{minipage} \qquad
\begin{minipage}{.4\textwidth}
  \centering
  \includegraphics[width=1\linewidth]{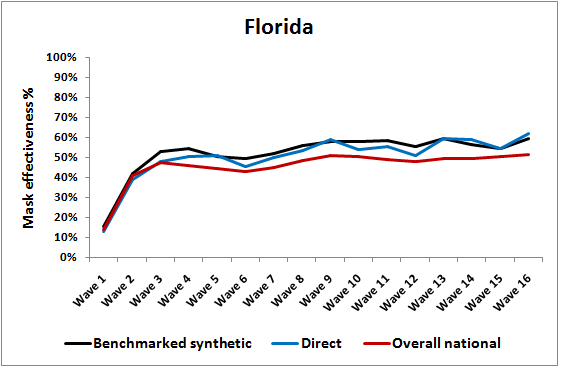}
  \end{minipage}
\vskip .2in
\caption{Time series trend of direct and benchmarked synthetic estimate for 6 sample states (3 small, 3 large)
}
\label{fig:eval_state_est}
\end{figure}

We now compare our synthetic estimates with the corresponding direct sample survey estimates (i.e., weighted proportion of people from UAS who believe mask is highly effective) for the 50 states and district of Columbia. This gives us an idea about the magnitude of bias in the synthetic estimates because direct estimates, though unreliable, are unbiased or approximately so. 
In Figure \ref{fig:eval_state_est}, we have 6 plots  corresponding to the same 6 states (3 with small population - District of Columbia, North Dakota, Rhode Island, and 3 with large population - California, New York, Florida) of point estimates (direct and benchmarked synthetic) vs waves, which display time series trends from wave 1 to wave 16. The direct estimates of mask effectiveness for small states could be unreasonable. For example, for DC direct estimates are 0\% for both waves 1 and 2.  On the other hand, benchmarked synthetic estimates 18\% and 39\% are more reasonable for these two waves  -- they are more in line with the national estimates for such waves. Similarly, for Rhode Island unreasonable 100\% mask effectiveness direct estimates for waves 13 and 15  have also been modified to a more reasonable benchmarked synthetic estimates.

\begin{figure}
\begin{center}
\begin{minipage}{.4\textwidth}
  \centering
  \includegraphics[width=1\linewidth]{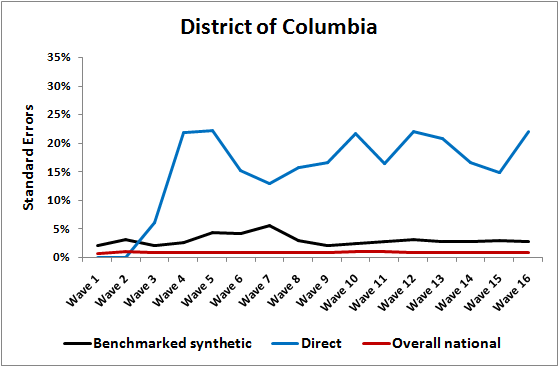}
  \end{minipage} \qquad
\begin{minipage}{.4\textwidth}
  \centering
  \includegraphics[width=1\linewidth]{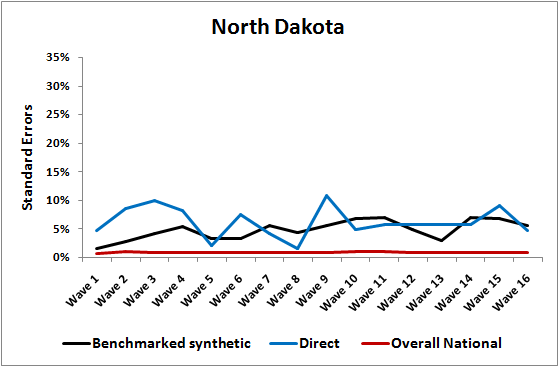}
  \end{minipage}
\vskip .2in
\begin{minipage}{.4\textwidth}
  \centering
  \includegraphics[width=1\linewidth]{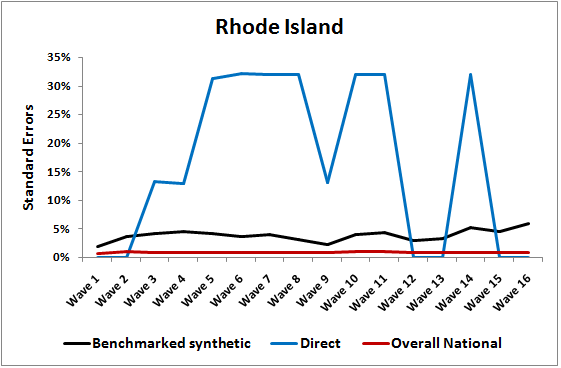}
  \end{minipage} \qquad
\begin{minipage}{.4\textwidth}
  \centering
  \includegraphics[width=1\linewidth]{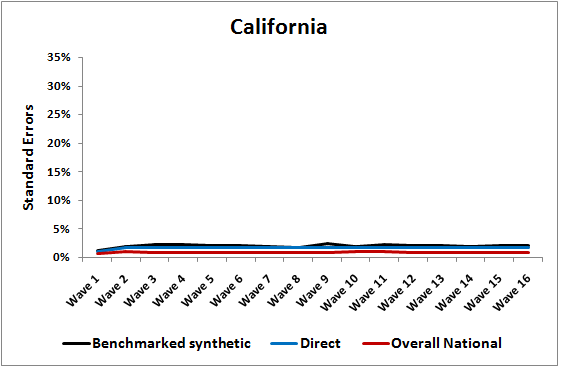}
 \end{minipage}
\vskip .2in
\begin{minipage}{.4\textwidth}
  \centering
  \includegraphics[width=1\linewidth]{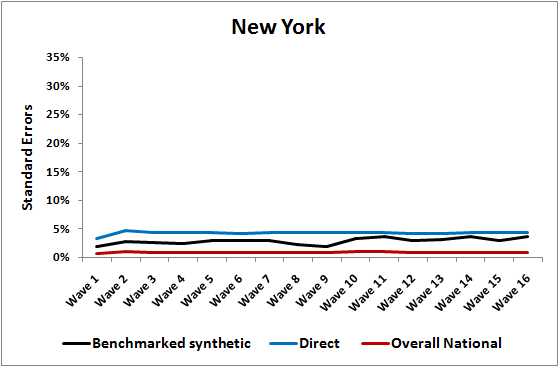}
  \end{minipage} \qquad
\begin{minipage}{.4\textwidth}
  \centering
  \includegraphics[width=1\linewidth]{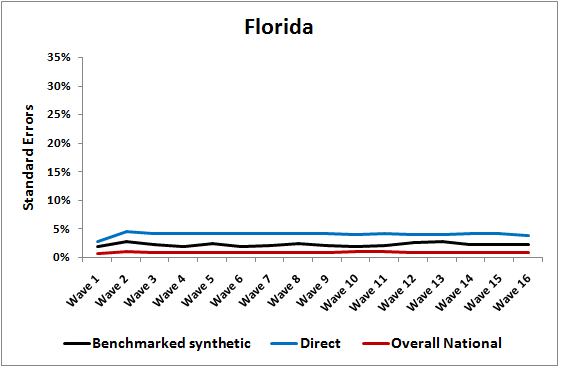}
 \end{minipage}
\end{center}
\vskip .2in
\caption{Time series trend of SE of direct and benchmarked synthetic estimate for 6 sample states (3 small, 3 large)
\label{fig:eval_state_SE}}
\end{figure}

Figure \ref{fig:eval_state_SE} displays SE of direct wave estimates and STD of  benchmarked synthetic estimate, computed using the proposed jackknife method.
If we focus on the error graphs the values from direct estimates get as high as 32\% for small states (i.e. one with low contribution to overall sample size). Using benchmarked synthetic estimates at state level, error has reduced to almost 6 times with as low as 6\% STD from jackknife method. For larger states like New York and Florida, the errors reduce using benchmarked synthetic estimate, although not to a great extent. For the state contributing most to the sample size, California, the STD and SE from jackknife and direct estimate are more or less similar. 

\begin{figure}[H]
    \begin{center}
    \includegraphics[width = 5in, height = 6cm]{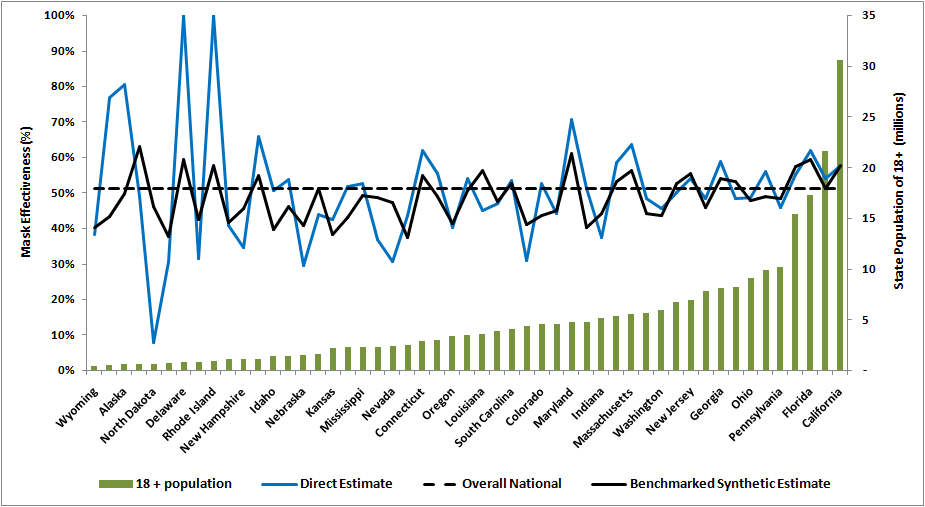}
    \end{center}
    \caption{State level comparison of direct estimator and synthetic estimator from M1
       on wave 16
    \label{fig: direct_vs_synthetic}}
    \end{figure}

 For the chosen model M2, we create a state level comparative diagram of benchmarked synthetic estimates with direct estimates in Figure \ref{fig: direct_vs_synthetic} with data from wave 16. As at state level the synthetic estimates and the corresponding benchmarked synthetic estimates  are really close, we have not plotted synthetic estimates for ease of viewing. We observe that our synthetic estimates are a much more stable one than the corresponding direct estimates. The states arranged in increasing order of total population show that the issue of highly variable state level direct estimates for the smaller states has been mitigated by the synthetic method. For largely populated states as well as for small ones the benchmarked synthetic estimates are doing good job of estimating the proportion of the response variable. We next check the robustness of the synthetic estimator in terms of variance through jackknife method.
   
\begin{figure}
\centering
\begin{minipage}{.4\textwidth}
  \centering
  \includegraphics[width=1\linewidth]{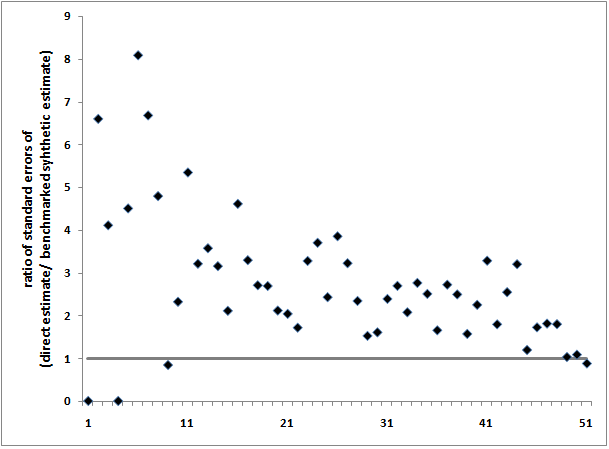}
 \end{minipage} \qquad
\begin{minipage}{.4\textwidth}
  \centering
    \includegraphics[width=1\linewidth]{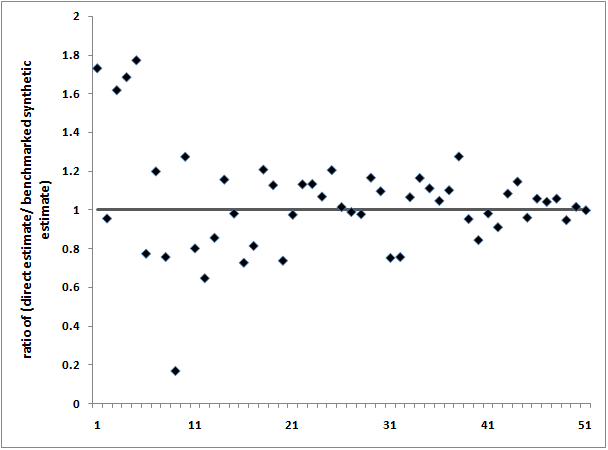}
\end{minipage}
\caption{Comparison of direct estimator with benchmarked synthetic estimator through ratio of SE/Jackknife STD and ratio of estimates from M1
on wave 16; domains arranged in increasing order of sample size.
\label{fig:ratio_estimate}}
\end{figure}

We fitted M2 on wave 16 and obtained jackknife estimates of variances and hence standard errors at state level.  We provide a comparative view with the SE from direct estimates at the state level. The two graphs in Figure \ref{fig:ratio_estimate} are based on the wave 16 data.  In the x-axis, states are arranged in increasing order of sample sizes.  In the first graph, the y-axis is the ratio of direct estimate (survey-weighted) and synthetic estimate.  In the second graph, the y-axis is the ratio of STD and SE, where SE is the standard error of  direct estimate coming right from UAS (treating states as domains) and STD is the jackknife standard error of  benchmarked synthetic estimate.  For states with small sample sizes (e.g. Rhode Island, Wyoming), we see a lot of differences between the survey-weighted direct estimates and the synthetic estimates. For states with large sample sizes (e.g., California), the ratio is approaching to 1 (as plotted by the straight line) as the auxiliary variables used to construct the synthetic estimator are reasonable.  We observe all the jackknife estimates are much smaller than direct estimates and we conclude that the model is a fair one at estimating the mask effectiveness at state level.
 
We define Benchmark Ratio (BR) as the ratio of the overall direct national estimate to the synthetic estimate (aggregated at the national level). The synthetic estimates, which are obtained at state level, are aggregated by multiplying by the ratio of the adult state population to the overall US adult population estimate and then adding up. The closer the value of BR is to 1 the better is the model. We see from Table \ref{tab: BR_details} that BR is close to 1 for all waves, using which we compute the Benchmarked or BR synthetic estimate.

    \begin{table}
    \caption{Benchmarking ratios and national synthetic and benchmarked synthetic estimates  for last five waves; synthetic estimates are based on Model 1.
    \label{tab: BR_details}}
    \begin{center}
    \begin{tabular}{rrrrrrr}
    \hline
    \hline
        Model & $0\%$  & $25\%$& $50\%$&$75\%$& $100\%$ & Mean \\
        \hline
        \shortstack{Benchmarking Ratio} & 0.98 & 0.98 & 0.99 & 0.99 & 1.00 & 0.99\\
        \shortstack{Synthetic estimate} & 14.08\% & 45.22\% & 48.63\% & 50.43\% & 51.94\% & 46.00\%\\
        \shortstack{BR Synthetic estimate} & 13.86\% & 44.62\% & 48.02\% & 49.66\% & 51.22\% & 45.38\%\\
        \hline
        \end{tabular}
        \end{center}
    \end{table}

\section{CONCLUSION}
The issue of mask effectiveness is a critical one and insight into this behavioral aspect of people helps in understanding the future impacts or spread of the disease at the state level. There is an impact of political affiliation and views on such behavioral aspect. Needless to say, it will be worthwhile to monitor this proportion in the coming time period after mid-November, which can be insightful as to the change in administration. From September we have seen a rise in cases in the US and in November we have seen total case count to have surpassed 11 million. Although there have been signs of downfall in the infection curve a new wave has arrived thereafter. Certainly this pandemic is still a serious health risk. Wearing masks is undoubtedly one of the few and most effective precautionary measures and people's awareness of such could be tracked from an analysis such as this one, which uses response of the public in surveys from a wide range of time period from March till November, 2020. Variation in perception through time is observed as the estimates show for overall US in March 14\% of people viewed mask wearing to be extremely effective, whereas in November 2020 this number is at around 51\%. 

The method of estimating population means or totals for the states of USA explained in the paper is a robust one which provides sensible and numerically sound estimates and the model selection and evaluation methods provide satisfactory results with all standard error of estimates within 2\% as per the standards. 
 We noticed high variability of synthetic estimates in state level estimation. We further note that while direct UAS estimates are designed to produce approximately unbiased estimates at the national level, they are subject to biases for the state level estimation.  Biases in the direct proportion estimates at the state level  may arise from the fact that they are essentially ratio estimates since the state sample sizes are random and expected sample sizes are small for most states.  Moreover, the UAS weights are not calibrated at the state level.

From our investigation, we found that synthetic estimates improve on UAS direct estimates in terms of variance reduction, especially for the small states. But since synthetic estimates are derived using a working model, they are subject to biases when working model is not reasonable. However, we observe that the benchmarking ratios for all waves are consistently around 1 showing lack of evidence for bias.  Our benchmarked synthetic estimates are close to the synthetic estimates because the benchmarking ratios are close to 1.  None-the-less by benchmarking synthetic estimates we achieve  data consistency and it is reasonable to expect to reduce biases as well.  We add that it is possible to reduce biases at the state level by benchmarking the synthetic estimates to the UAS direct estimates for a goup of states (e.g., benchmarking with a division).  This may be needed for other synthetic estimation problems.

This method can be replicated or tried out for any binary variable as has been done for mask effectiveness and even for categorical or continuous ones. There are numerous interesting areas in the UAS survey, which can be studied to find state level estimates. To name a few areas: physical health, mental health, economic and financial anxiety, etc.

\section{ACKNOWLEDGEMENTS}
The project described in this paper relies on data from survey(s) administered by the Understanding America Study, which is maintained by the Center for Economic and Social Research (CESR) at the University of Southern California. The content of this paper is solely the responsibility of the authors and does not necessarily represent the official views of USC or UAS. The collection of the UAS COVID-19 tracking data is supported in part by the Bill \& Melinda Gates Foundation and by grant U01AG054580 from the National Institute on Aging.

\bibliographystyle{chicago}
\bibliography{references}

\end{document}